\documentclass[runningheads]{llncs}
\usepackage[utf8]{inputenc}
\usepackage{multirow} 
\usepackage{graphicx}
\usepackage{array}
\usepackage{amsmath}
\usepackage{amsfonts}
\usepackage{amssymb}
\usepackage{bm}
\usepackage{subcaption}

\usepackage{soul}
\usepackage{amsmath}
\usepackage{amssymb}
\usepackage[table]{xcolor}

\usepackage{anyfontsize}
\begin{document}

\title{Few-Shot Domain Adaptive Object Detection for Microscopic Images}
%
%

\author{Sumayya Inayat, Nimra Dilawar, Waqas Sultani, Mohsen Ali}
\authorrunning{S. Inayat et al.}
%
 \institute{Intelligent Machines Lab, Department of Artificial Intelligence, \\
 Information Technology University, Pakistan\\
 {\fontsize{7}{6}\selectfont  \{sumayya.inayat, phdcs21003, waqas.sultani, mohsen.ali\}@itu.edu.pk}} 

%
\maketitle              

\begin{abstract}
 {
In recent years numerous domain adaptive strategies have been proposed to help deep learning models overcome the challenges posed by the domain shift. 
However these even unsupervised domain adaptive strategies still require large amount of the target data. 
On the other hand medical imaging datasets are often characterized by class imbalance, scarcity of both labeled and unlabeled data. 
Few-shot domain adaptive object detection (FSDAOD) addresses the challenge of adapting object detectors to target domains with limited labeled data. 
However, existing works are not successful in dealing with randomly selected target domain images which may not accurately represent the real population, and result in overfitting to small validation sets and poor generalization to larger test sets.} 
Medical datasets often exhibit high-class imbalance and background similarity, leading to increased false positives and lower mean Average Precision (mAP) in target domains. 
To overcome these challenges, we propose a novel FSDAOD strategy for microscopic imaging. Our contributions include:  a domain adaptive class balancing strategy for few shot scenario; multi-layer instance-level inter and intra-domain alignment by enhancing similarity between the instances of classes regardless of the domain and enhance dissimilarity when it's not.
Furthermore, an instance-level classification loss is applied in the middle layers of the object detector to enforce the retention of features necessary for the correct classification regardless of the domain. 
Extensive experimental results with competitive baselines indicate the effectiveness of our proposed approach by achieving state-of-the-art results on two public microscopic datasets.
{\fontsize{7.5}{6.5}\selectfont https://github.com/intelligentMachinesLab/few-shot-domain-adaptive-microscopy.} 

\end{abstract}
\section{Introduction}
Deep-learning-based approaches have played a crucial role in microscopic cell detection \cite{thomas2017review,yang2017faster,huang20212,fujita2020cell,zhao2021positive} but these approaches require abundant expert annotated data which is very difficult to obtain due to the experts' time and availability. 
Alternatively, Few-Shot Object Detection (FSOD) methods come into play where most of the works utilize the available dense-labeled dataset (base set) to train a base model and then utilize the few-labeled image samples (support set), coming from the same image distribution but different label space, to fine-tune the model.  {However, in real world it is not always viable to get the base set and support set from same data distribution. The few image samples might come from a different distribution but with similar label space, a challenge known as Few-Shot Domain Adaptive Object Detection (FSDAOD) \cite{gao2022acrofod,gao2023asyfod}. The discrepancy in the source (abundant train data) domain and the target (few-shot samples) domain \cite{zhang2023spectral} arises due to differences in the data acquisition protocols, including factors such as microscopic quality, lighting conditions, microscopic-lens resolution, camera-lens quality and so forth.} It becomes further challenging due to the extreme data imbalance \cite{gosain2017handling} in the microscopic cells, (especially in Few-shot system), categories, and the visual similarity between the background and foreground, as well as the intra-foreground visual similarity, leading to a higher false-positive and false negative rate.

Existing methods \cite{wang2020frustratingly,han2023few,wang2024fine,wu2020meta} solve the FSOD challenge by employing meta-learning-based approaches, where they pre-train a model with base (abundant) classes and fine-tune over novel (scarce) classes coming from similar distribution. 
These approaches fail to work when training and testing data are from different distributions. To tackle this limitation, recently unsupervised and few-shot domain adaptive approaches are proposed \cite{zhang2023spectral,gao2022acrofod,gao2023asyfod}. Although useful, however as pointed out by our experimental results, these methods suffer from over-fitting to the smaller validation set and weak generalization to large test sets.

Since in few-shot domain adaptive object detection, only few samples of any class are available from the target domain, a strategy for the feature alignment could be to enforce that representation of same class across domain is same. 
However, for this strategy to work, the representation of samples from same class in the soure domain should also be very similar. To address this challenge, we propose Intra-Inter-Domain Feature Alignment technique; {I2DA}, that addresses 
\textbf{(a)} the domain shift between similar class cells by aligning the inter-domain feature level representations of cells coming from same classes, and  \textbf{(b)} Intra-Domain Feature Consistency at the cell level to learn distinguishable features for each class because the foreground cells in microscopic datasets possess high visual similarity with the background cells. This is especially challenging in the case of malarial-affected cells where for example, `Ring' class is very similar to the background platelets, resulting in a higher rate of false positives leading to a lower mAP \%. Secondly, we propose a Domain-Generalized Class Balancing Cut-Paste strategy; {CBCP} to tackle the extreme class imbalance in microscopic datasets, which balances the overall count of the rare and abundant classes in the data. We strategies that class imbalance should be handled by generating samples of rare classes through selected visual augmentation of its existing samples. Further, we cut-paste these generated samples by carefully selecting locations in the images where no other cells are present. 
The extensive experimentation on two public microscopic datasets M5-Malaria \cite{sultani2022towards} and Raabin-WBC \cite{kouzehkanan2022large} demonstrate the effectiveness of our method by outperforming with an increase in average mAP@50 by 
8.3 point and 
14.6 points (respectively) as compared to other competitive baselines.

\begin{figure}[t]
  \begin{minipage}[t]{\linewidth}
    \centering
    \includegraphics[width=0.9\linewidth]{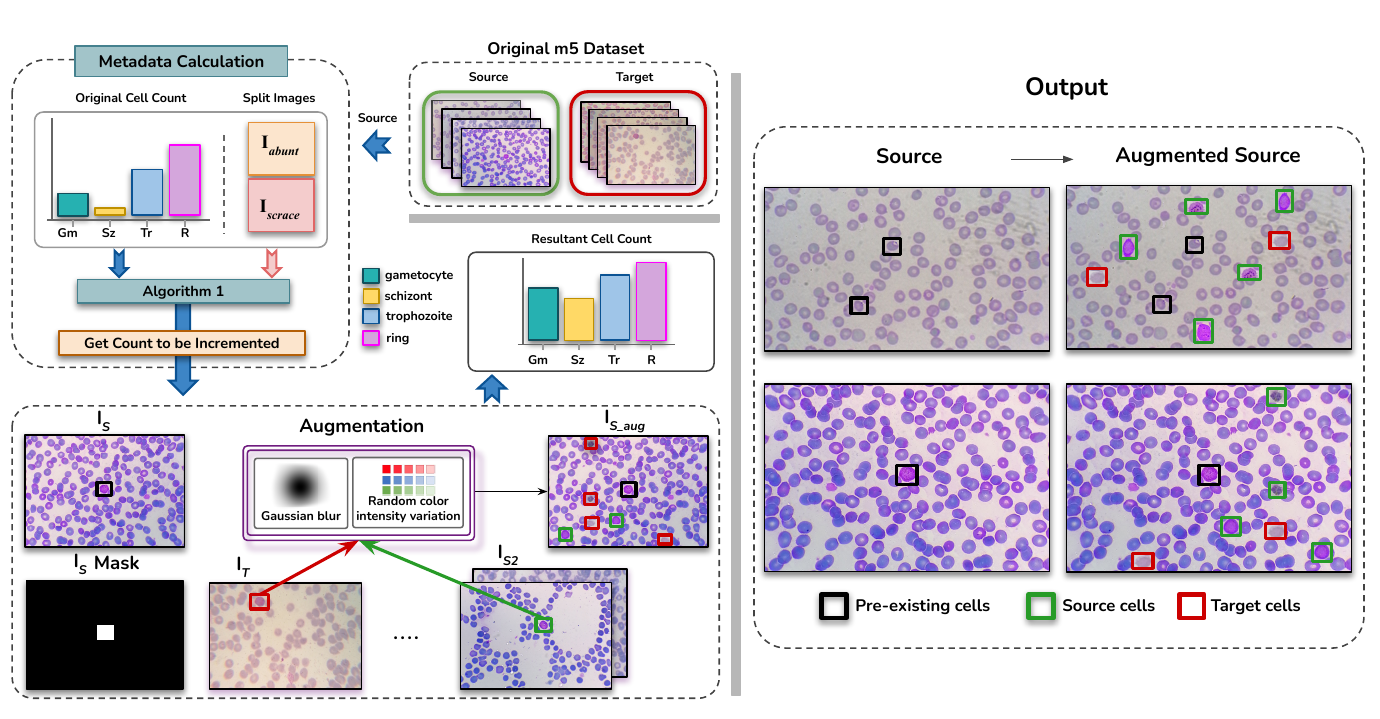}
    \caption{Class Balancing Cut Paste strategy: We first compute the metadata and increment-stats from the abundant source dataset and few target images and then increment cells to the images with less pre-existing cells.} 
    \label{fig:cut-paste-flow}
  \end{minipage}
\end{figure}

\section{Methodology}


\noindent\textbf{Preliminaries:} Assume we have access to two datasets, a source domain dataset \bm{$D_s$} and target domain few-shot dataset \bm{$D_t$}. Note that, \bm{$|D_s|$ $>>$ $|D_t|$}. 
The \bm{$N$} be the total number of classes, and both datasets suffer from class imbalance. 
Objective is to train the object detector $\mathcal{F}$, using the large dataset \bm{$D_s$} and few shot dataset \bm{$D_t$} such that it generalizes to the target domain. 
\subsection{Class Balancing Domain Generalized Cut-Paste}

To address the limitations of existing image resampling techniques \cite{lin2014microsoft,deng2009imagenet}, we propose to increase the count of rare class instances to \"match\ the maximum instances class count. Instead of making it exact same, we rather make the sizes comparable, since that is more realistic.

\noindent\textbf{Class Balancing Cut Paste Strategy}:
Given, \bm{$D_s$} and \bm{$D_t$},
we construct a new dataset \bm{$D_{aug}$} from their combination,  such that \bm{$|D_{aug}|=|D_{s}|$} but has balanced count (Fig \ref{fig:cut-paste-flow}). We analyze \bm{$D_s$} for data statistics,
(1) the total object count of each class \bm{$C$},(2) images \bm{$I_{sc}$} with less than a threshold \textbf{$r$} amount of pre-existing objects per image, (3) \bm{$I_{mr}$} images with greater than threshold $r$ amount of pre-existing objects per image, (4) \bm{$P$} images with \bm{$i^{th}$} class present. 
We only increment cells to \bm{$I_{sc}$} to avoid hard augmentation and ensure more realistic real world simulation. Next, we compute the increment stats, (Algorithm in supplementary), for \bm{$i^{th}$} class having total object count less than the max object count. The stats determine, the total number of times each instance of the \bm{$i^{th}$} class (a) has to be incremented, (b) the \bm{$i_{sc} \subset I_{sc}$} images it has to be incremented in, (c) and the times it will be incremented in one \bm{$i_{sc}$} image. Further, we take \bm{$i_{sc}$}, and generate a binary mask \bm{$B$} associated with the objects in \bm{$i_{sc}$}. Followed by extracting an \bm{$i^{th}$} class object \bm{$O$} from \bm{$p^{th}$} image in \bm{$P$} as per the increment stats, apply random visual augmentations and find the empty region \bm{$er$} of pasting in \bm{$i_{sc}$}. We find \bm{$er$} in \bm{$B$}, by choosing a slot that has \bm{$iou=0$} with any pre-existing objects in \bm{$i_{sc}$} and paste \bm{$O$} in place of \bm{$er$}. To achieve a domain generalized \bm{$i_{sc}$}, we extract a random cell from a random image from \bm{$D_t$} and paste it in each \bm{$i_{sc}$} likewise and update the corresponding annotation file of \bm{$i_{sc}$} for all the incremented objects. For random visual augmentation, we choose random color intensity variation and random Gaussian blurring (see supplementary). 

We denote the new object wise augmented images as \bm{$I_{aug}$}. The final per-class count in \bm{$I_{aug}$} is not supposed to be perfectly equal because we don't want it to deviate from the real-world scenarios. Finally the resultant dataset $D_{aug}$ is a combination of \bm{$I_{aug}$} and \bm{$I_{mr}$}. The overall flow of the process is shown in Fig \ref{fig:cut-paste-flow}. 
\begin{figure}[t]
  \begin{minipage}[t]{\linewidth}
    \centering
    \includegraphics[width=0.9\linewidth]{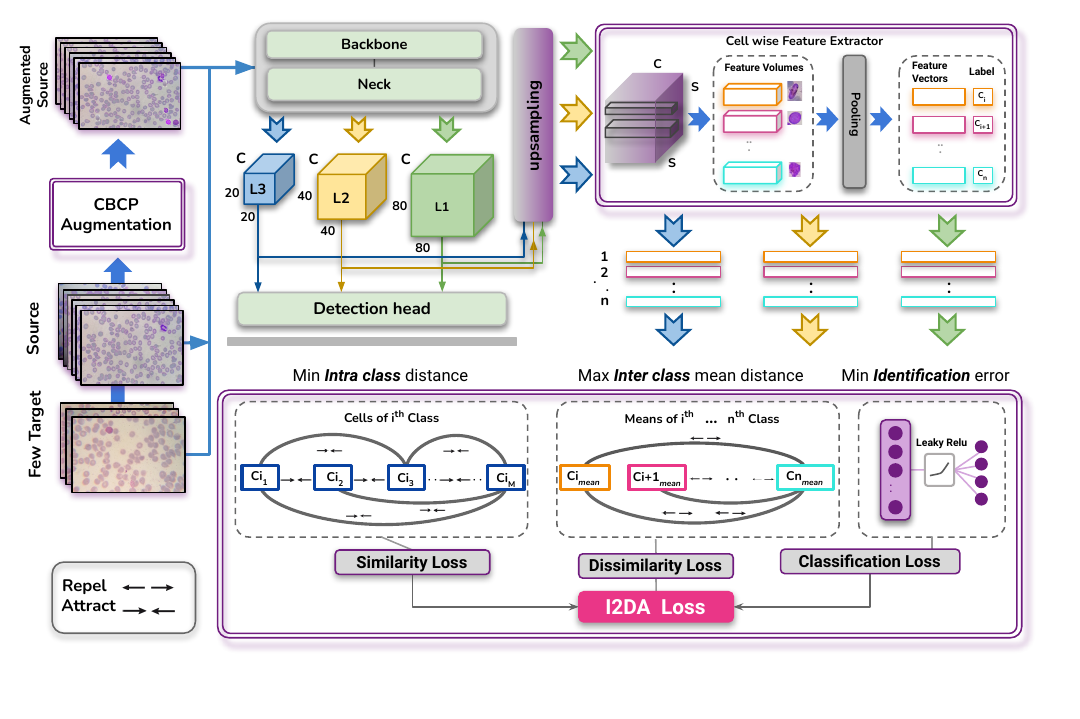}
    \caption{Proposed approach: We first build our class-wise balanced dataset through a cut-paste strategy (Fig \ref{fig:cut-paste-flow}), then train the model with our proposed inter-domain instance feature-level alignment and intra-domain instance feature-level consistency. We extract multi-layer neck features and upsample them to a common size, followed by the extraction of pooled object-level features, which are then passed to the similarity-dissimilarity and classification module.
    } 
    \label{fig:workflow}
  \end{minipage}
\end{figure} 
\subsection{Inter-domain Alignment and Intra-domain Class Consistency}
To align features in space across the domains, it is required to maximize the dissimilarity between the instances of different classes and minimize the similarity between instances of similar classes. Traditionally object level contrastive loss \cite{cao2023contrastive} is used for this purpose but we argue that only contrastive loss is not enough for robust feature-level alignment especially in a challenging dataset like Malaria which has a high foreground-background visual similarity that leads to increased false positive rate. To ensure full feature level alignment, one must compute the intra-domain feature similarity and dissimilarity as well and not just inter-domain. Intra-domain is essential to minimize the false positive and false negative rate. Furthermore, an additional feature-level instance classification guidance is required to boost the model performance.


Therefore to achieve our purpose we formulate a novel solution, illustrated in Fig \ref{fig:workflow}, aiming to learn more robust instance features. More specifically we design a module that computes the similarity between instances of similar classes and computes dissimilarity between the instances of dissimilar classes. We endorse that similar class feature-level representation must be near in feature space and dissimilar class features must be distant in space irrespective of their domains. The similarity or dissimilarity must be performed within the domain as well to learn the diversity of intra-class features. We boost the learning with a classifier to learn more robust features for each class. We extract multi-layer instance-level features from the neck of the detector $\mathcal{F}$. The reason behind choosing multi-layer neck features is to take the most representative small, medium, and large-size object-level features. We up-sample these features to ($S\times S$) and extract all the cell features corresponding to its ground truth, followed by average pooling. Next, we compute $c$ pairwise combinations of all the $m$ instances belonging to $i^{th}$ class, and compute their mean similarity loss $l_{c\_mean}$. 
\begin{table}[t]
\fontsize{7.85pt}{12.5pt}\selectfont

\caption{Results in mAP@50(\%) on Malaria\cite{sultani2022towards} test set.}

\label{tab:malrai_results_few_shot}
\begin{tabular}{>{\raggedright\arraybackslash}m{1.56cm} |c c c|c c c|c c c|c c c|c c c}
\hline
\multicolumn{16}{c}{Malaria-HCM-1000x $ \rightarrow $ Malaria-LCM-1000x} \\
\hline
Method &
\multicolumn{3}{c}{mAP@50(\%)} &
\multicolumn{3}{c}{Gametocyte} &
\multicolumn{3}{c}{Schizont} &
\multicolumn{3}{c}{Trophozoit} &
\multicolumn{3}{c}{Ring}\\
\hline
Source & 
\multicolumn{3}{c}{19.9} &
\multicolumn{3}{c}{3.9} &
\multicolumn{3}{c}{0.5} &
\multicolumn{3}{c}{55.9} &
\multicolumn{3}{c}{19.3}\\

Oracle & 
\multicolumn{3}{c}{43.7} &
\multicolumn{3}{c}{33.3} &
\multicolumn{3}{c}{4.3} &
\multicolumn{3}{c}{81.6} &
\multicolumn{3}{c}{55.7}\\

\hline
 Shots & 2 & 3 & 5 & 2 & 3 & 5 & 2 & 3 & 5 & 2 & 3 & 5 & 2 & 3 & 5\\

\hline
FsDet\cite{wang2020frustratingly} & 11.2 & 11.5& 12.9 & 11.8 &11.2& 12.3 & 0.0 & 0.0 & 0.0 & 27.7 & 28.3& 30.7 & 5.4 & 6.5 & 8.6  \\
VFA \cite{han2023few}& 8.9 & 6.5 & 8.1 & 9.1 & 3.2 & 24.5 & 0.1 & 3.3 & 0.2 & 17.8 & 19.4 & 6.5 & 12.8 & 0.2  & 1.3\\
FDP\cite{wang2024fine} & 14.6 & 14.7 & 20.5 & 16.4 & 8.1 & 33.7 & 1.20 & 10.1 & 1.3 & 27.4 & 24.4 & 33.3 & 13.5 & 16.4 & 14.0\\

AsyFOD\cite{gao2023asyfod} & 26.0 & 29.1 & 33.5 & 14.9 & 23.3 & 36.8 & 1.20 & 7.0 & 2.80 & \underline{59.4} & 60.7 & \underline{64.7} & \underline{28.7} & 31.8 & \underline{30.9}\\


AcroFOD\cite{gao2022acrofod} & \underline{32.9} & \underline{42.5} & \underline{39.1} & \underline{27.6} & \underline{50.9} & \underline{62.9} & \textbf{17.6} & \textbf{22.1}  & \underline{5.40} & 58.7  & \underline{62.7} & 61.3 & 27.8 & 
\textbf{34.4} & 27.0\\
\hline




Ours & \textbf{44.7}& \textbf{45.9} & \textbf{48.9} & \textbf{71.4} & \textbf{66.0} & \textbf{68.2} & \underline{11.4} & \underline{18.1} & \textbf{30.4}  & \textbf{66.9} & \textbf{66.7} & \textbf{65.6} & \textbf{29.3} & \underline{32.6} & \textbf{31.5} \\
\hline
\end{tabular}
\end{table}

Specifically, let $\mathbf{v}_{ik}$ and $\mathbf{v}_{il}$ denote the feature vector $k$ and $l$ in $i^{th}$ class, then their cosine similarity can be denote by $\text{sim}(\mathbf{v}_{ik}, \mathbf{v}_{il})$. We sum up all the $l_{c\_mean}$ losses to compute the overall similarity loss $L_{sim}$:

\begin{equation}
L_{\text{sim}} = \sum_{i} \frac{1}{{\binom{n_i}{2}}} \sum_{k=1}^{n_i-1} \sum_{l=k+1}^{n_i} \text{sim}(\mathbf{v}_{ik}, \mathbf{v}_{il})
\end{equation}



Next, the dissimilarity loss is computed, for which we first compute $N$ class mean feature vectors $\mathbf{\bar{v}_N}$. Each $\mathbf{\bar{v}_i}$ is a mean of the $n$ instances of $i^{th}$ class. Followed by computing $\mathbf{\bar{c}}$ pairwise combinations of $\mathbf{\bar{v}_N}$ and calculating the dissimilarity $d$ between the $\mathbf{\bar{v}}_k$ and $\mathbf{\bar{v}}_l$, where $k, l$ are mean features of 2 different classes. $d$ is computed using cosine similarity $s_{d}$ with margin = $m$. If $s_{d} < m$, set $d=0$. Finally, the total dissimilarity loss $L_{dis}$ can be obtained by summing up the resulting $d$ values for all pairwise combinations of mean feature vectors:

\begin{equation}
L_{\text{dis}} = \sum_{k=1}^{N-1} \sum_{l=k+1}^{N} \max\left(0, \left( \frac{\mathbf{\bar{v}}_k \cdot \mathbf{\bar{v}}_l}{\|\mathbf{\bar{v}}_k\| \|\mathbf{\bar{v}}_l\|}\right) - m\right)
\end{equation}

\begin{table}[t]
\fontsize{7.85pt}{12.5pt}\selectfont
\caption{Results in mAP@50(\%) on Raabin-WBC \cite{kouzehkanan2022large} test set.} 
\label{tab:raabin_results_few_shot}
\begin{tabular}{>{\raggedright \arraybackslash} m{1.68cm} |c c c|c c c|c c c|c c c|c c c}
\hline
\multicolumn{16}{c}{Raabin-WBC-HCM $ \rightarrow $ Raabin-WBC-LCM} \\
\hline
Method  &
\multicolumn{3}{c} {mAP50(\%)}&
\multicolumn{3}{c} {Large Lymph} &
\multicolumn{3}{c} {Neutrophil} &
\multicolumn{3}{c} {Small Lymph} &
\multicolumn{3}{c} {Monocyte} \\

\hline
Source &
\multicolumn{3}{c}{27.2} &
\multicolumn{3}{c}{25.1} &
\multicolumn{3}{c}{59.6} &
\multicolumn{3}{c}{22.9} &
\multicolumn{3}{c}{1.0}\\

Oracle &
\multicolumn{3}{c}{75.0} &
\multicolumn{3}{c}{90.9} &
\multicolumn{3}{c}{98.1} &
\multicolumn{3}{c}{83.2} &
\multicolumn{3}{c}{27.7}\\
\hline
Shots & 2 & 3 & 5 & 2 & 3 & 5 & 2 & 3 & 5 & 2 & 3 & 5 & 2 & 3 & 5\\

\hline
FsDet\cite{wang2020frustratingly}
 & 26.5 & 28.5 & 30.1 & 41.5 & 24.3 & 38.1 & 17.5 & 31.8 & 44.1 & 23.7 & 34.4 & 29.6 & \underline{23.1} & \underline{23.7} & 8.7 \\


VFA\cite{han2023few}
& 30.3 & 33.2 & 45.2 & 44.6  & 28.6 & 59.8 & 25.7 & 59.8 & 66.0  & \underline{48.7} & 27.7 & \underline{42.6} & 2.40 & 16.7 & 12.5 \\
FDP\cite{wang2024fine}
 & 35.9 & 32.4 & 44.3 & 28.2 & 36.2 & 59.1 & 60.7 & 39.8 & 66.8 & 46.2 & 34.0 & 35.2 & 8.60 & 19.9 & \underline{15.9}\\
 
 
AsyFOD\cite{gao2023asyfod}
& 35.6 & 38.1 & 26.3 & 37.2  & 42.2 & 39.3 & 58.3 & 48.7 & 43.1  & 46.5 & \textbf{51.1} & 22.4 & 0.3 & 10.4  & 0.3\\


AcroFOD \cite{gao2022acrofod}& \underline{44.9} & \underline{47.2} & \underline{61.2} & \underline{50.5}  & \underline{64.1} & \textbf{82.1} & \textbf{88.1} & \textbf{89.1} & \textbf{95.9} & 37.6 & 30.7 & \textbf{59.6} & 3.5 & 5.1 & 7.3 \\
\hline




Ours  &  \textbf{64.2}  &\textbf{62.6}  &\textbf{70.8} & \textbf{74.1}  & \textbf{76.0} & \underline{75.2} & \underline{87.2} & \underline{86.0} &  \underline{94.3} & \textbf{54.7}  & \underline{46.8} & 42.5 &\textbf{40.6} & \textbf{41.6}  & \textbf{71.3} \\
\hline
\end{tabular}
\end{table}

Further, we compute the $N$ class-wise classification losses, let $l_{ik}$ denote the class loss of instance $k$ in $i^{th}$ class, then take the mean of all $l_{i}$ in class $i$, and add up all such mean losses for each class  Let $m$ be the number of instances in class $i$, then the instance level classification loss $L_{cls}$ is given by:
\begin{equation}
L_{\text{cls}} = \sum_{i=1}^{N} \left( \frac{1}{n_i}\sum_{k=1}^{n_i} l_{ik} \right)
\end{equation}

Finally, we compute the mean similarity, mean dissimilarity, and mean classification losses for the three levels, followed by multiplication with threshold $\lambda1$, $\lambda2$ and $\lambda3$ with the similarity, dissimilarity, and class mean losses respectively. We add up the losses as our final $I2DA$ loss $L_{I2DA}$ is represented as:
\begin{equation}
\text{$L_{I2DA}$} = \lambda_1L_{sim} + \lambda_2L_{dis} + \lambda_3L_{cls}
\end{equation}


\section{Experiments and Results}
\noindent\textbf{Datasets:}
\textbf{M5} \cite{sultani2022towards} is a large-scale malarial domain adaptive cell detection dataset captured from two different microscopes, one high cost, and one low cost, and the corresponding images captured from three different resolution levels. We utilized their standard train val test splits for training whereas for few-shot we randomly sampled a set of 8 images as per 
\cite{gao2023asyfod,gao2022acrofod} while also selected images as per\cite{han2023few,wang2019few,wang2024fine,wu2020meta} for 2-shot, 3-shot, and 5-shot. We consider the shots as the number of images per a specific category.
\noindent \textbf{Raabin-WBC} is a white blood cell dataset where  11000 images were taken high-cost microscope data  and 4000 were taken using low-cost microscope. The authors did not provide any standard train,val, or test splits for the detection task, hence we first extracted center cropped images and as per 'Label2', selected images for the four following classes, Large Lymph, Neutrophill, Small Lymph, and Monocyte. We then made equal random splits of the train, val, and test for both the microscope data and chose the few-shot samples similar to M5.

\begin{table}[t]
\fontsize{8pt}{12pt}\selectfont
\caption{mAP@50(\%) on \cite{sultani2022towards} $\&$ \cite{kouzehkanan2022large} test sets on 8 random few-target images.}
\label{tab:results-8-few}
\begin{tabular}{lcccccccccc}
\hline
\multicolumn{1}{l|}{\multirow{2}{*}{\begin{tabular}[c]{@{}l@{}}Data \\ \end{tabular}}} & \multicolumn{5}{c|}{Malaria}                                          & \multicolumn{5}{c}{Raabin-WBC}                       \\ \cline{2-11} 
\multicolumn{1}{l|}{}                                                                      & mAP@50    & Gamet.    & Schizo.    & Troph.    & \multicolumn{1}{c|}{Ring}          & mAP@50   & L-Lymp.    & Neutro.      & S-Lymp.    & Mono.         \\ \hline

\multicolumn{1}{l|}{AsyFOD}                                                                & 30.2          & 23.8           & 1.3 & \underline{61.8}          & \multicolumn{1}{c|}{\textbf{33.9}} & 33.7          & 28.4          & 48.6          & \textbf{56.2}          & 1.4           \\


\multicolumn{1}{l|}{AcroFOD}&\underline{33.1} & \underline{46.8} & \textbf{3.8}    &56.9 & \multicolumn{1}{c|}{24.9}& \underline{48.9}   &\underline{69.1}  &\textbf{90.7}  & 27.7 & \underline{7.9} \\
\hline




\multicolumn{1}{l|}{Ours }                                                                 & \textbf{40.3} & \textbf{62.3} & \underline{2.0}            & \textbf{64.4}    & \multicolumn{1}{c|}{\underline{32.6}}    & \textbf{55.7}& \textbf{71.3}          & \underline{80.6}          & \underline{49.6}            & \textbf{23.9} \\ \hline
\end{tabular}
\end{table}


\noindent\textbf{Implementation Details:}
Our techniques are object detector agnostic, however, for experiments we have used \cite{glenn_jocher_2020_4154370} as base model. 
For our experiments (conducted on GTX1080 GPU), batch-size was set to 4.
We develop customized batches for each epoch such that each batch of the extracted features contains $n \geq 1$ object from the few-shot target set 
For each batch we select 2\% of the batch size from  target, 30\% real source, and 68\% augmented source data set.
The $\lambda$1 $\lambda$2 and $\lambda$3 values are set to 0.005, 0.005, and 0.001 respectively.

\subsection{Results}
We perform two sets of experiments, one with 8 random images as per \cite{gao2023asyfod,gao2022acrofod} that may have any number of images per class or even miss a rare class. 
The other set with the k-shot settings \cite{wang2020frustratingly,han2023few,wang2024fine}. We define our shots as k images per class. Malaria results of the baselines and our work on 2-shot, 3-shot, and 5-shot images are shown in Table \ref{tab:malrai_results_few_shot}. We evaluated our models based on mAP@50 because all the given baselines \cite{wang2020frustratingly,han2023few,wang2024fine,gao2023asyfod,gao2022acrofod} yielded results in mAP@50. But for \cite{gao2023asyfod,gao2022acrofod} we also evaluated the models on mAP@50:95 and average precision and average recall as well. Please refer to the supplementary for detailed results. As shown in Table \ref{tab:malrai_results_few_shot}, 
our work over-performs the existing competitive baseline by a good margin. The reason is clear enough because our methodology does not over-adapt to the small few-image set and performs a more robust feature-level alignment. Recall-comparison graph, in supplementary, shows the improved recall of our proposed method as compared to \cite{gao2023asyfod,gao2022acrofod} 
because their work is less optimized for small-sized cells whereas our multi-level feature alignment in our work ensures various sized objects learned properly.
Table \ref{tab:raabin_results_few_shot} shows the results of the Raabin-WBC test set and proves that our method works well for large-size objects as well. The respective recall comparison (see suplemetary) also supports the highest recall of Raabin-WBC cells by our method. Further results can be found in the supplementary material. Fig \ref{fig:qual-resulst} shows the qualitative results of AcroFOD \cite{gao2022acrofod} and our methods. AcroFOD has given some wrong predictions and was comparatively less confident in the correct predictions. In contrast, ours is more confident in the correct predictions and the false-positive rate is comparatively less. Table \ref{tab:results-8-few} shows our results obtained on 8 random target domain images and as visible our method overperforms in these settings as well.

\begin{figure}[t]
  \begin{minipage}[t]{\linewidth}
    \centering
    \includegraphics[width=0.8\linewidth]{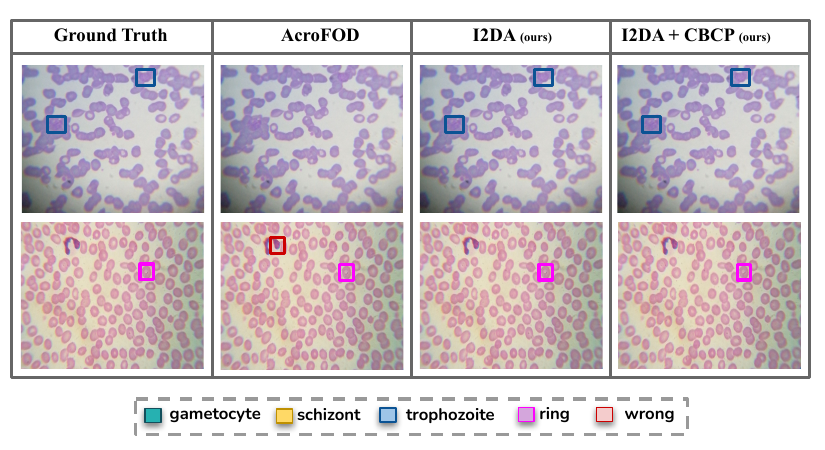}
    \caption{Qualitative results of LCM Malarial defected regions after adaptation. 
    }
    \label{fig:qual-resulst}
  \end{minipage}
\end{figure}

\section{Conclusion}
We have provided a novel solution to tackle \textbf{FSDAOD} in few shot settings in microscopic imaging. The intra-class feature space variation is minimized and inter-class variation is maximized irrespective of domains which further boosted the performance with a specialized feature-level instance classifier. To handle the extreme class imbalance in microscopic datasets especially in domain adaptive few-shot settings, we devise a novel strategy to balance the skewed data distribution with our cut-paste augmentation strategy. Extensive experimentation validate the effectiveness of our method as compared to the existing competitive baselines. Our method achieved an increase of average 
8.3 points in mAP@50 even with 2-shot settings on Malaria \cite{sultani2022towards} datasets, validating its effectiveness on medium to small sized cells. Whereas we achieve an increase of average 
14.7 points in mAP@50 for Raabin-WBC dataset that has big to medium sized cells. We further look forward to extending our augmentation strategy to real world scenarios.

%
%
\bibliographystyle{splncs04}
\bibliography{paper}
%






\end{document}